\documentstyle[prb,aps,amstex,graphics,twoside,twocolumn,multicol]{revtex}
\newcommand{\bra}[1]{{\langle\,#1\,|}}
\newcommand{\ket}[1]{{|\,#1\,\rangle}}
\begin{document}
\title{Dynamical properties of a two--dimensional electron gas in a magnetic field within the composite fermion model} 
\author{Darren J.T. Leonard and Neil F. Johnson} 
\address{Department of Physics, Clarendon Laboratory, Oxford University, Parks Road, 
Oxford OX1 3PU, UK}
\maketitle
\begin{abstract} 
We investigate the response of a two--dimensional electron gas, in the fractional quantum Hall regime, to the 
sudden appearance of a localised charged probe using the Chern--Simons theory of composite fermions.
The dynamic structure factor of the electron gas is found to 
have a major influence on the spectral function of the probe. In particular, there is an orthogonality catastrophe 
when the filling factor is an even--denominator filling fraction due to the compressibility of the state, but 
there is no catastrophe at odd--denominator filling factors because these states have a gap to excitations. The catastrophe
is found to be more severe for composite fermions in zero \emph{effective} magnetic field than it is for electrons in zero 
\emph{real} magnetic field.
Oscillations in the spectral function, arising when the composite fermions are at integer filling, have a period equal to the 
composite fermion cyclotron energy. We propose a tunneling experiment which directly measures the spectral function  
from which one could determine the composite fermion effective mass. 
\end{abstract}
\bigskip
\begin{center}
\end{center}
\section{Introduction}
The properties of the two--dimensional electron gas (2DEG), subject to a strong magnetic field applied perpendicular 
to the plane of the 2DEG, may be interpreted in terms of particles similar to electrons which experience a weak effective 
magnetic field--they interact with each other not only through the Coulomb interaction but also via a Chern--Simons potential
which couples the particle density to the particle current. The Chern--Simons theory of these composite fermions 
(CFs)~\cite{Jain,Zhang,HLR,Lopez1,Lopez2,Stern1,Stern2,Simon1,Simon2,Simon3,Simon4}
has been used extensively to understand the properties of the 2DEG in the fractional quantum Hall regime (FQHR). In particular, the
fractional quantum Hall effect exhibited by the electron gas is understood to be a manifestation of the underlying integer
quantum Hall effect of the CF gas; the relation between the filling factor of the electron system $\nu$
and that of the CF system $\nu^\star$ is given by
\begin{equation}
\nu=\frac{\nu^\star}{2m\nu^\star\pm1}.
\end{equation}
At filling factors of the form $\nu=1/2m$, where $m$ is an integer, the 
CFs experience zero effective magnetic field and the CF excitation 
spectrum is gapless: the 2DEG is compressible. Various experiments have been performed on the 2DEG in the FQHR which 
seem to be in agreement with the CF theory,~\cite{HLR} such as those experiments involving surface acoustic 
waves,~\cite{HLR,HalperinB,Willett1,Willett2} magnetotransport measurements~\cite{Du1,Du2,Leadley1,Leadley2,Manoharan}
and experiments in which electrons tunnel into the 2DEG.~\cite{Eisenstein,Chang}
\par
In this paper, we examine the properties of the 2DEG within the CF model in order to calculate the 
response of the CF system to the sudden appearance of a localized charge. We show how this charge may be used to probe
the charge--density excitation spectrum of the 2DEG, and propose an experiment which obtains this information directly.
\par
The outline of the paper is as follows. In Sec.~\ref{Sect:QDGF} we show how the sudden appearance of the probe charge near to the 
2DEG affects the spectral function of the probe. We derive the dependence of the spectral function on the excitation spectrum of the
2DEG, i.e. the dynamic structure factor. In particular, we show that the probe spectral function will depend on whether or not 
there is a gap to excitations: the spectral function may exhibit an orthogonality catastrophe if the system is gapless, but will
not do so when there is a gap. 
\par
Section~\ref{Sect:EDS} calculates the dynamic structure factor $S({\mathbf{q}},\omega)$ 
for the 2DEG at even--denominator filling factors using the theory of CFs in zero effective field; the modified random phase 
approximation (MRPA) is used. We compare this with a similar calculation for electrons in zero real field, and find that the 
CF system in zero effective field has a much higher number of low energy excitations than does the electron gas at zero real field. 
We then derive the fluctuations in the charge density induced by the sudden appearance of the probe charge and find that the 
relaxation time of the fluctuations is
much greater in the CF gas than in the electron gas because of the larger number of excitations. Finally in this section, we
derive the spectral function of the probe charge along the lines of Sec.~\ref{Sect:QDGF} utilizing the calculated
dynamic structure factors. The CF system is shown to exhibit an orthogonality catastrophe which is more severe than that of 
the electron gas: the reason for this is related to the number of low--energy excitations in each system.
\par
In Sec.~\ref{Sect:ODS}, the calculations from Sec.~\ref{Sect:EDS} are 
generalized to CFs in finite effective fields $\nu^\star=1, 3,
7$ and 10 which map onto the 2DEG at $\nu=1/3$, 3/7, 7/15 and 10/21. There is a gap to excitations, therefore there
is no orthogonality catastrophe. We find that the probe spectral function has oscillations, the envelope of which 
approaches the shape of the spectral function at $\nu=1/2$ as the effective field tends to zero. Moreover, these oscillations
have a period equal to the effective cyclotron energy at the appropriate filling factor. 
\par
In Sec.~\ref{Sect:Tunneling} we propose an experiment in which a 2DEG in the FQHR is probed by the sudden appearance 
of an electron in a quantum dot situated a distance $d$ above the plane of the 2DEG. Electrons in a source lead are allowed
to tunnel resonantly into a quantum dot which is kept at a fixed energy; this dot acts as a monochromator. 
The electron is then allowed to tunnel into a second quantum dot which is situated a small distance $d$ above the plane of the 2DEG;
the energy of this probe dot relative to the monochromator dot is controlled by a gate voltage. 
As soon as the electron appears in the dot, excitations are created in the 2DEG to screen the probe; the energy required by the electron 
to do this is manifest as broadening of the dot spectral function to higher energies. The current is proportional to the density of 
states at the energy of the monochromator dot, hence measuring the current as a function of the voltage difference 
between the two dots
will give the spectral function of the probe dot coupled to the 2DEG. As the results in Secs.~\ref{Sect:EDS}~and~\ref{Sect:ODS}
imply, one should be able to detect the presence of compressible states at even--denominator filling factors as an orthogonality 
catastrophe in the I--V charactersitic, and one may determine a composite fermion effective mass at odd--denominator filling 
factors. Section~\ref{Sect:Conclusions} contains our conclusions.
\section{Spectral function of the quantum dot}\label{Sect:QDGF}
The quantum dot is assumed to have an unperturbed energy $\epsilon_{\text{QD}}$ and interacts with the CFs only when it is 
present. Since there cannot be more than one quantum dot, the quantum dot operators $d$ and $d^\dag$ 
are fermionic, i.e. they anticommute. The Hamiltonian of the whole system is given by
\begin{equation}\label{eq:qdhamiltonian}
H=H_0 + \epsilon_{\text{QD}} d^\dag d + V d^\dag d
\end{equation}  
where $H_0$ is the full CF Hamiltonian, and
\begin{equation}
V=\sum_{\mathbf{q}}\,V({\mathbf{q}})\,\hat{\rho}({\mathbf{q}})
\end{equation}
is the interaction of the CFs with the quantum dot. 
$\hat{\rho}({\mathbf{q}})$
is the CF density operator. In terms of the bare Coulomb potential
$V_0({\mathbf{q}})=e^2/2\epsilon_0 \epsilon_r q $
\begin{equation}
V({\mathbf{q}})= V_0({\mathbf{q}}) f_{\text{QD}}({\mathbf{q}})
f_{\text{z}}({\mathbf{q}})
\end{equation}
where the form factor, dependent on the in--plane wavefunction $\psi_{\text{QD}}({\mathbf{r}})$ 
of the quantum dot , is given by
\begin{equation}
f_{\text{QD}}({\mathbf{q}})=\int\,d^2 r 
|\psi_{\text{QD}}({\mathbf{r}})|^2 e^{i{\mathbf{q}}\cdot{\mathbf{r}}}.
\end{equation}
The form factor for the direction perpendicular to the plane is
\begin{equation}
f_{\text{z}}({\mathbf{q}})=\iint\,dz\,dz' |\psi_{\text{2DEG}}
(z)|^2 e^{-q|z-z'|} |\psi_{\text{QD}}(z')|^2
\end{equation}
where $\psi_{\text{2DEG}}(z)$ and the $\psi_{\text{QD}}(z')$
are the wavefunctions for the 2DEG and the quantum dot respectively in the perpendicular
direction. In this article the 2DEG is assumed to have zero thickness for simplicity. 
The quantum dot wavefunction is detailed in those sections where a specific form is required, 
i.e. in Sections~\ref{Sect:results:chargedensity}~and~\ref{Sect:tunnel:expt}.
\par
One may define two sets of eigenstates for the 2DEG. The  states $|n\rangle$ are eigenstates of the 
unperturbed 2DEG Hamiltonian $H_0$, while the tilded states 
 $|\tilde{n}\rangle$ are the eigenstates of the perturbed 2DEG Hamiltonian
$\tilde{H}_0=H_0+V$. Therefore
\begin{equation}
H_0|n\rangle = \epsilon_n|n\rangle\hspace{2em}\tilde{H}_0|\tilde{n}\rangle = 
\epsilon_{\tilde{n}}|\tilde{n}\rangle.
\end{equation}
The time--ordered quantum dot Green's function $G_{\text{QD}}(t-t')$ is defined via 
\begin{equation}\label{eq:GQD1}
i\hbar\,G_{\text{QD}}(t-t')\equiv\,\langle T [d(t)\,d^\dag(t')]\rangle.
\end{equation}
Therefore the spectral function of the quantum dot is
\begin{multline}\label{eq:overlap}
A_{\text{QD}}(\omega)=-2\,\text{Im}\int^\infty_{-\infty}\hspace{-0.5em}dt\,e^{i\omega t}\,G_{\text{QD}}(t)\\
=\frac{2\pi}{\hbar}\,\sum_{\tilde{n}}\,|\langle\,0\,|\,\tilde{n}\,
\rangle|^2\,\delta(\omega-\omega_{\text{QD}}+\omega_0-\omega_{\tilde{n}}).
\end{multline}
This equation relates the quantum dot spectral function directly to the
overlap of the ground state of the unperturbed 2DEG and the states of the
perturbed 2DEG. The spectral function at $\omega=\omega_{\tilde{n}}-\omega_{\tilde{0}}
+\omega_{\text{QD}}$ is in fact proportional to $|\langle\,0\,|\,\tilde{n}\,\rangle|^2$. Since $|\,\tilde{0}\,\rangle$ is the
lowest--energy perturbed 2DEG eigenstate the spectral function has
a threshold frequency of $\omega=\omega_{\tilde{0}}-\omega_{0}+\omega_{\text{QD}}$
below which it is identically zero. The difference
$\omega_{\tilde{0}}-\omega_{0}$ can also be thought of as a renormalisation of the
quantum dot energy by writing 
$\bar{\omega}_{\text{QD}}=\omega_{\tilde{0}}-\omega_{0}+\omega_{\text{QD}}$.
The threshold therefore occurs when $\omega=\bar{\omega}_{\text{QD}}$.
This relation between the spectral function of the quantum dot and 
the overlap of the unperturbed ground state and the perturbed excited
states, is very powerful. It relates directly to the 
so-called Anderson orthogonality 
catastrophe.~\cite{Mahan,Anderson,Nozieres,Langreth,Shung,Ohtaka,Uenoyama2,Westfahl}
This was formulated for the conduction band electrons in a metal, and says that the 
ground state of a system of $N$ electrons is orthogonal to the
ground state of the $N$--electron system in the presence of a static potential.
Therefore, in this case the quantum dot spectral function must vanish at
threshold precisely because $\langle\,\tilde{0}\,|\,0\,\rangle=0$. This has very
important consequences when
probing systems by introducing a localized potential. Indeed, this is precisely
what happens in the X-ray edge problem giving rise to 
Fermi-edge singularities.~\cite{Mahan,Nozieres}
\par
\onecolumn
\begin{multicols}{2}
Having identified the formal properties of the Green's function, in a real calculation
one usually starts from a position where the perturbed (i.e. tilded)
eigenstates are not known. Thus one must perform perturbation theory in the potential $V$
in order to determine the quantum dot spectral function.
The Green's function~(\ref{eq:GQD1}) may be written as~\cite{Mahan,Nozieres,Negele}
\begin{gather}
\begin{split}
i\hbar G_{\text{QD}}(t-t')&=\\\theta(t-t')&\,\exp[-i\omega_{\text{QD}}(t-t')]
\,\langle\,0\,|\,S(t,t')\,|\,0\,\rangle\\
\end{split}\label{eq:GQD2}\\
S(t,t')\equiv T\,\exp\biggl[-\frac{i}{\hbar}
\int^t_{t'} \hspace{-0.5em}dt_1\,V(t_1) \biggr]
\end{gather}
with $V(\tau)\equiv \exp(iH_0 \tau/\hbar)\,V\,\exp(-iH_0 \tau/\hbar)$.
Equation~(\ref{eq:GQD2}) may be rewritten in terms of 
an exponential resummation (the cumulant expansion)~\cite{Mahan,Negele} as
\begin{multline}\label{eq:GQD3}
i\hbar G_{\text{QD}}(t-t')=\theta(t-t')\,\\\exp[-i\omega_{\text{QD}}(t-t')]
\exp \biggl[\sum_{n=1}\,F_n(t-t')\biggr]
\end{multline}
where
\begin{multline}\label{eq:Fterms}
F_n(\tau)=\fracwithdelims(){-i}{\hbar}^n\,\frac{1}{n}\times\\\int_0^\tau \hspace{-0.5em} dt_1
\ldots \int_0^\tau \hspace{-0.5em}dt_n  \langle\,0\,|\,T\,V(t_1)\ldots V(t_n)\,|\,0\,\rangle
_{\text{connected}}\ .
\end{multline}
$T$ is the time--ordering operator, and the subscript `connected'
indicates one should calculate only the connected diagrams in an expansion of the Green's functions.~\cite{Mahan,Negele}
Thus far, no approximation has been made. However, it is not possible to calculate all of 
the terms in Eq.~(\ref{eq:Fterms}) up to infinite order for a general system. The exception is the case of non--interacting 
particles where the most divergent (logarithmic) terms may be summed to all 
orders.~\cite{Mahan,Nozieres,Ohtaka,Matveev} However, the interactions can prove to be 
extremely important in determining the shape of the quantum dot spectral function.~\cite{Mahan,Langreth,Shung} 
Assuming a weak potential, which is the case in this article, the most important contributions arise from the first
few terms, i.e. $F_1(\tau)\text{ and }F_2(\tau)$. 
One finds that for $\tau>0$
\begin{gather}
F_1(\tau)=0\\
F_2(\tau)=i\tau\Delta-\Gamma(\tau)
\end{gather}
where
\begin{gather}
\Delta=\int_0^\infty\hspace{-0.5em}d\omega\, \frac{\rho(\omega)}{\omega}\\
\Gamma(\tau)=\int_0^\infty \hspace{-0.5em}d\omega\, \frac{\rho(\omega)}{\omega^2}(1-e^{-i\omega\tau})\\
\rho(\omega)=\frac{1}{\hbar} \sum_{\mathbf{q}} |V({\mathbf{q}})|^2 S({\mathbf{q}},\omega).
\end{gather}
These expressions have the following physical interpretation~\cite{Mahan}. 
$S({\mathbf{q}},\omega)$ is the dynamic structure of the 2DEG, which gives the density of pair excitations of 
momentum ${\mathbf{q}}\text{ and energy }\omega$. The quantum dot then
couples to these pair excitations with a strength $|V({\mathbf{q}})|^2$ creating a
density of pair excitations $\rho(\omega)$ at energy $\omega$.
The mean total energy $\Delta$ of the excitations created is an energy renormalisation of the quantum dot Green's function,
while $L=\int_0^\infty d\omega\,\frac{\rho(\omega)}{\omega^2}$ is the mean total number of excitations created.~\cite{Mahan} 
Consequently, the quantum dot spectral function is
\begin{equation}\label{eq:spectral0}
A_{\text{QD}}(\omega)=\frac{2}{\hbar}\text{Re}\,\int_0^\infty \hspace{-0.5em}dt\,
e^{i(\omega-\bar{\omega}_{\text{QD}})t-\Gamma(t)},
\end{equation}
where the renormalised quantum dot energy $\bar{\omega}_{\text{QD}}=\omega_{\text{QD}}-\Delta$.
The interpretation of Eq.~(\ref{eq:spectral0})
is that in order to put the quantum dot in its state, one has to supply it with extra energy which is  
used to create the excitations of the system which screen the quantum dot. However, if one did not know about the coupling
of the quantum dot to the system, one would just assume that the quantum dot had a broadened density of states.~\cite{Mahan} 
Expanding the factor $\exp-\Gamma(t)$ as a power series shows how the density of pair excitation $\rho(\omega)$ contributes
to the shape of the spectral function $A_{\text{QD}}(\omega)$; the total excitation at a given energy $\omega$ is in fact the sum of
all possible combinations of pair excitations with total energy $\omega$.
The spectral function in Eq.~(\ref{eq:spectral0}) may be written
\begin{multline}\label{eq:spectral2}
A_{\text{QD}}(\omega)=\frac{2}{\hbar}\exp(-L)\times \\\text{Re}\,\int_0^\infty\hspace{-0.5em}dt\,
e^{i(\omega-\bar{\omega}_{\text{QD}})t} \exp\biggl[\int_0^\infty\hspace{-0.5em}
d\omega'\,\frac{\rho(\omega')}{\omega^{'  2}}\,e^{-i\omega' t} \biggr]
\end{multline}
Note that the Debye--Waller factor $\exp(-L)$~\cite{Mahan} may formally be taken outside of the integral. 
Expanding the exponential in the usual way, one finds for $\epsilon\geq 0$
\end{multicols}
\begin{equation}\label{eq:specexpansion}
\begin{split}
A&_{\text{QD}}(\epsilon+\bar{\omega}_{\text{QD}})=\frac{2\pi}{\hbar}\,
\exp(-L)\,\text{Re}\,\int_0^\infty \hspace{-0.5em}dt\,e^{i\epsilon t} \biggl[1+\sum_{n=1} \frac{1}{n!}
\int_0^\infty\hspace{-0.5em}d\omega_1 \dots \int_0^\infty\hspace{-0.5em}d\omega_n\, 
e^{-it(\omega_1+\dots+\omega_n)} \frac{\rho(\omega_1)}{\omega^2_1}
\ldots \frac{\rho(\omega_n)}{\omega^2_n} \biggr]\\
&=\frac{2\pi}{\hbar}\,\exp(-L)\,\biggl[\delta(\epsilon)
+\frac{\rho(\epsilon)}{\epsilon^2}+\frac{1}{2!}\,\int_0^\epsilon\hspace{-0.5em} d\omega_1\,
\frac{\rho(\omega_1)\,\rho(\epsilon-\omega_1)}{\omega^2_1\,(\epsilon-\omega_1)^2}
+\frac{1}{3!}\,\int_0^\epsilon\hspace{-0.5em}d\omega_1 \int_0^{\epsilon-\omega_1}\hspace{-2em}d\omega_2\,
\frac{\rho(\omega_1)\,\rho(\omega_2)\,\rho(\epsilon-\omega_1-\omega_2)}
{\omega^2_1\,\omega^2_2\,(\epsilon-\omega_1-\omega_2)^2} +\ldots\biggr].\\
\end{split}
\end{equation}
\begin{multicols}{2}
The first term in Eq.~(\ref{eq:specexpansion}) is the contribution from
the creation of zero pairs, and is just a $\delta$--function at the 
threshold $\epsilon=0$. The second term is the contribution from
a single pair excitation of energy $\epsilon$. The third part gives 
the contribution of a two--pair excitation with total energy $\epsilon$.
One pair has energy $\omega_1$ while the other has  $\epsilon-\omega_1$,
and we must integrate over all $\omega_1$ to count all possible pairings.
\par
There is however an important point to be made regarding Eq.~(\ref{eq:specexpansion}).
An orthogonality catastrophe will arise in a system if the mean number of excitations $L$ diverges~\cite{Mahan} because 
\begin{equation}
\lim_{L\rightarrow \infty} \exp(-L)=0
\end{equation}
i.e. the Debye--Waller factor is zero. In this case therefore, it is not possible to separate out $\Gamma(t)$ as is done in 
Eq.~(\ref{eq:spectral2}), nor is it possibe to generate the expansion in Eq.~(\ref{eq:specexpansion}). 
If $L$ diverges, the weight of the $\delta$--peak at $\epsilon=0$ is zero and the quantum dot spectral function
at $\omega=\bar{\omega}_{\text{QD}}$ is identically zero. 
However, Eq.~(\ref{eq:overlap}) says that the quantum dot
spectral function at this frequency is proportional to $|\bra{\tilde{0}}\,0\,\rangle|^2$, thus this must also
be zero. Hence, the two states are orthogonal and there will be an orthogonality catastrophe. 
The criterion for $L$ not to have an infra-red divergence is either that the system has a gap to excitations, 
in which case the convergence is trivial, or that the low energy behaviour of the density of pair excitations has a stronger
than linear dependence. For low $\omega,\ \rho(\omega)\text{ typically goes as }
\omega^{1-\eta}$; there is usually a cut-off $\epsilon_0$ in the excitation 
spectrum which, for electrons, is of order the Fermi energy.~\cite{Mahan} Thus there is no ultraviolet divergence. 
Hence the mean number of excitations becomes
\begin{equation}
\begin{split}
L&=\int_0^\infty\hspace{-0.5em}d\omega\,\frac{\rho(\omega)}{\omega^2}\\
&=\lim_{\epsilon \rightarrow 0} \int_\epsilon^{\epsilon_0}\hspace{-0.5em}d\omega\,\omega^{-1-\eta}\\
&=\lim_{\epsilon \rightarrow 0}
\begin{cases}
-\ln(\epsilon)& \text{if } \eta=0\\
\frac{\epsilon^{-\eta}}{\eta}& \text{otherwise}
\end{cases}\ .
\end{split}
\end{equation} 
So it is easily seen that if $\eta\geq 0$ then $L$ diverges, and
there is an orthogonality catastrophe. Otherwise, there is no 
catastrophe. If this is the case, a $\delta$--function occurs at 
threshold corresponding to creation of zero--pair excitations. The weight
is however reduced by the factor $\exp(-L)$ because the coupling
to the system means that $L$ is non-zero. The spectrum close to 
threshold is itself basically the first term in the expansion. The higher
order contributions from multipair excitations have a much smaller weight;
however, these terms play a greater role further from threshold. 
\section{Even--denominator states}\label{Sect:EDS}
In the CF picture of the states in the FQHR with $\nu=1/2m$, the system is viewed as a 
Fermi--liquid of CFs with an effective mass $m^\star\propto\sqrt{B}$, which
experience an effective magnetic field $B^\star=0$.~\cite{HLR,HalperinB}
Hence the single--particle states are plane waves with a momentum
 ${\mathbf{K}}$ and energy $\epsilon=\hbar^2 k^2 / 2m^\star$. The density of CFs is the 
same as that of the electrons, $n_{\text{e}}$, thus the states fill up to the Fermi energy
 $\epsilon_f\equiv\hbar\omega_f=\hbar^2 k^2_f / 2m^\star$ with the Fermi wave vector $k_f=\sqrt{4\pi n_{\text{e}}}$.
\subsection{The dynamic structure factor}
The neutral density excitations of the Fermi surface are quasi--particle quasi--hole pairs with 
energy $\hbar\omega$ and momentum ${{\mathbf{q}}}$, the spectrum of which is contained in the 
the dynamic structure factor
\begin{equation}
S({\mathbf{q}},\omega)=-\frac{1}{\pi e^2 c^2} \text{Im} K_{00}({\mathbf{q}},\omega)
\end{equation}
where ${\mathbf{K}}=(\mathbf{1}-{\mathbf{K}}^0\bar{{\mathbf{D}}}^0)^{-1}{\mathbf{K}}^0$. 
Here, ${\mathbf{K}}$ is the retarded electromagnetic response kernel which is calculated in the modified random phase approximation
(MRPA). In the MRPA, the CFs interact with each other not only through the Coulomb $V({\mathbf{q}})$ and Chern--Simons 
potentials contained in the gauge--field propagator 
\begin{equation}
{\mathbf{D}}^0({\mathbf{q}},\omega)=
\begin{pmatrix} \dfrac{V({\mathbf{q}})}{e^2 c^2} & -\dfrac{i2m\phi_0}{ec|{\mathbf{q}}|}\\
\ & \ \\
\dfrac{i2m\phi_0}{ec|{\mathbf{q}}| } & 0\\
\end{pmatrix}
\end{equation}
but also through a Landau interaction term 
\begin{equation}
\boldsymbol{\Lambda}({\mathbf{q}},\omega)=\frac{m_{\text{b}}\alpha_1}{e^2 n_{\text{e}}} \begin{pmatrix} 
\dfrac{\omega^2}{c^2 q^2} & \ 0\ \\ 
\ & \ \\
\ 0\ & \ 1\ \\ \end{pmatrix}
\end{equation}
which ensures Galilean invariance and satisfaction of Kohn's theorem. Here, $\alpha_1$ relates the effective mass $m^\star$ and
the band mass $m_{\text{b}}$ by $m^\star=m_{\text{b}} (1+\alpha_1)$. Thus 
$\bar{{\mathbf{D}}}_0({\mathbf{q}},\omega)={\mathbf{D}}({\mathbf{q}},\omega)+\boldsymbol{\Lambda}({\mathbf{q}},\omega)$.
The bare retarded electromagnetic response kernel ${\mathbf{K}}^0({\mathbf{q}},\omega)$ has components
\begin{gather}
K^0_{00}({\mathbf{q}},\omega)\equiv\frac{e^2 c^2 m^\star}{4\pi\hbar^2} \Pi_0(x,y)\\
K^0_{11}({\mathbf{q}},\omega)\equiv\frac{e^2 n_{\text{e}}}{12 m^\star} \Pi_1(x,y)\\
K^0_{01}({\mathbf{q}},\omega)=K^0_{10}({\mathbf{q}},\omega)=0
\end{gather}
where $x=q/k_f$ and $y=\omega/\omega_f$, and
\begin{multline}
\Pi_0(x,y)=\frac{1}{x} \Bigl[ -2x - \text{sgn}(X)\,\theta[|X|-2]\sqrt{X^2-4}\\
+\text{sgn}(Y)\,\theta[|Y|-2]\sqrt{Y^2-4}\\
+i\bigl(\theta[2-|Y|]\sqrt{4-Y^2}\\
-\theta[2-|X|]\sqrt{4-X^2}\bigr)\Bigr]
\end{multline}
\begin{multline}
\Pi_1(x,y)=\frac{1}{x} \Bigl[Y^3-\text{sgn}(Y)\,\theta[|Y|-2]\sqrt{(Y^2-4)^3}\\
-X^3-\text{sgn}(X)\,\theta[|X|-2]\sqrt{(X^2-4)^3}\\
+i\bigl(\theta[2-|Y|] \sqrt{(4-Y^2)^3}\\
-\theta[2-|X|] \sqrt{(4-X^2)^3}\bigr) \Bigr]
\end{multline}
with
\begin{gather}
X=\frac{y}{x}-x\\
\intertext{and}
Y=\frac{y}{x}+x.
\end{gather}
The dynamic structure factor may thus be written
\begin{gather}
S({\mathbf{q}},\omega)=\frac{m_{\text{b}}}{\pi \hbar^2}\,\sigma(x,y)\\
\sigma(x,y)=-\frac{m^\star}{4\pi m_{\text{b}}}\, \text{Im} \Pi(x,y)
\end{gather}
where the dimensionless density--density correlation function is found to be
\end{multicols} 
\begin{equation}\label{eq:pidefined}
\Pi(x,y)=\frac{\Pi_0\biggl(1+\dfrac{\Gamma}{12}\Pi_1\biggr)}
{\biggl(1-\dfrac{5\sqrt{m}\Pi_0}{3x}+\dfrac{\Gamma y^2 \Pi_0}{4 x^2}\biggr)\biggl(1+\dfrac{\Gamma}{12}\Pi_1\biggr)
-\dfrac{m^2\Pi_0 \Pi_1}{12 x^2}}
\end{equation}
\begin{multicols}{2}
\noindent with 
\begin{equation}
\Gamma=\frac{m_{\text{b}}-m^\star}{m^\star}.
\end{equation}
Equation~(\ref{eq:pidefined}) uses the effective mass
\begin{equation}
m^\star=\frac{4\pi\epsilon_0 \epsilon_{\text{r}} \hbar^2}{C e^2 l_{\text{c}}},
\end{equation}
which scales as the square--root of the magnetic field. The coefficient $C$ is determined by
equating the energy gaps determined by the CF theory for some of the Jain states around 
half--filling with those obtained in the exact diagonalisation of small systems on a sphere.~\cite{HLR,dAmbrumenil} 
The value determined by HLR is $C\approx 0.3$. For a 2DEG with $n_{\text{e}}=10^{15}$ m$^{-2}$ and with 
$\epsilon_{\text{r}}=13$ this implies an effective mass at half--filling $m^\star=0.26m_{\text{e}}$.
The actual masses measured in experiments are likely to depend on the amount of Landau level mixing and 
on the finite thickness of the 2DEG, both of which tend to increase the mass.~\cite{HLR,Chakraborty} 
Therefore the HLR mass is expected to be a lower limit for the true effective mass. 
\subsection{Pair continuum}
The dynamic structure is easily evaluated numerically straight from the MRPA expression 
for the response function. Figure~\ref{fig:struc1/2} shows the dimensionless quantity $\sigma(x,y)$ 
for electrons at a filling factor $\nu=1/2$, i.e. the CFs experience an effective magnetic field $B^\star=0$.
Figure~\ref{fig:strucelec} gives the structure factor for electrons at $B=0$ which have a mass equal to that of 
CFs for $\nu=1/2$. These electrons are essentially the same as the CFs in zero effective field except they interact only
through the Coulomb potential and not via the off--diagonal Chern--Simons terms. 
There are two clear features in each of Figs.~\ref{fig:struc1/2}~and~\ref{fig:strucelec}. 
The first is the dominant central structure which is 
the contribution from the pair continuum. Since the calculation is essentially of an RPA type, it includes
only single pair excitations.~\cite{Pines} An excitation created with a momentum ${\mathbf{q}}$ involves the transition
of one particle from an occupied to an unoccupied state. The RPA or MRPA structure factor is obtained by
allowing for the interactions between this particle--hole pair. Multipair excitations with momentum
${\mathbf{q}}$ involve simultaneous transitions of more than one particle from occupied to unoccupied states,
which are not accounted for in the RPA. Consequently, the region of the pair continuum has the same bounds 
as it would for non--interacting particles, namely for a given momentum $x$ there will be excitations
with an energy $y$ such that $(x+1)^2-1\leq y \leq (x-1)^2-1$.
\par
The most important difference between Figs.~\ref{fig:struc1/2}~and~\ref{fig:strucelec} 
is that, at low energies and momentum, $S(\mathbf{q},\omega)$ is much larger
in the CF system at $B^\star=0$ than it is for the electron system at $B=0$. Since the only 
difference between the CFs and the electrons is that the CFs interact
via the Chern--Simons potential as well as the Coulomb potential, the greater number of low energy 
and momentum excitations is attributable to scattering by the Chern--Simons field. As a result, the 
screening properties of the CF liquid are quite different from those of the electron liquid,
and this has a marked effect on the response to applied potentials, as will be shown later.
\par
The second feature in Figs.~\ref{fig:struc1/2}~and~\ref{fig:strucelec} is the black line in 
the $\sigma=0$ plane which appears to the right of the pair continuum. This is the collective mode,
which for the CFs is the cyclotron mode, and for the electrons is the plasmon mode. The 
contribution to the structure factor is not plotted, since it is a delta--function singularity along the 
mode with a weight many times larger than that of the pair continuum. 
\subsection{The cyclotron mode, the diffusive mode and the f--sum rule}
The collective mode is identified as the locus of points (x,y) for which the denominator of the
density--density correlation function $\Pi(x,y)$ vanishes.~\cite{Mahan,Negele,Pines} For a given momentum, the pole 
in frequency is found to lie on the real axis. However, because of causality the poles are 
pushed just slightly below the real axis. The structure factor coming from this mode is found to be 
\begin{equation}
\sigma(x,y)=\frac{m^\star}{4\pi m_{\text{b}}} \text{Res}\, \Pi(x,y)\,\delta(y-y_{\text{c}}(x))
\end{equation}
where Res$\,\Pi(x,y)$ is the residue of $\Pi$ and the mode has dispersion $y_{\text{c}}(x)$.
Expanding the denominator of $\Pi(x,y)$ for low $x$ but finite $y$ leads to the following 
equations for the collective mode
\begin{equation}
\begin{split}
y=&\frac{2m}{1+\Gamma}-\frac{m\Gamma x^2}{6(1+\Gamma)^2} \text{ for CFs} \\
y^2=&\frac{20\sqrt{m} x}{3(1+\Gamma)} \text{ for electrons}.\\
\end{split}
\end{equation}
Thus Kohn's theorem is obeyed for CFs because, in the limit $x\rightarrow 0$, the 
pole occurs at the cyclotron energy
\begin{equation}
\omega=\frac{2m\omega_{\text{f}}}{1+\Gamma}=\omega_{\text{c}}.
\end{equation}
The plasmon mode for the electrons at $B=0$ is found to be gapless and varies as the square--root of the momentum,
as it should in two dimensions.~\cite{Mahan}
\par
An expansion of the denominator of $\Pi(x,y)$ for low $y/x$ and $x$ reveals that the 
CF spectrum also has a diffusive mode, i.e. the pole in the complex frequency plane
lies on the imaginary axis.~\cite{HLR} In fact, in this regime one finds that
\begin{equation}\label{eq:diffmode}
\Pi(x,y)\approx\frac{-x}{\dfrac{5\sqrt{m}}3-\dfrac{i m^2y}{x^2}}
\end{equation}
which clearly has a pole at $y=-i\zeta x^2$, where $\zeta=5/(3m\sqrt{m})$. This is a diffusive mode,
since it is the dispersion relation for normal modes of the diffusion equation
\begin{equation}
D\nabla^2\rho({\mathbf{r}},t)=\frac{\partial \rho}{\partial t}({\mathbf{r}},t).
\end{equation} 
\par
One of the consequences of the equation of continuity is that the dynamic structure factor must 
satisfy the f--sum rule~\cite{Pines} 
\begin{equation}
\int^\infty_0 \hspace{-0.5em}d\omega\,\omega\,S({\mathbf{q}},\omega) = \frac{n_{\text{e}} 
{\mathbf{q}}^2}{2m_{\text{b}}},
\end{equation}
and it was found that the sum rule is satisfied to within a fraction of one percent.
This is as expected since the MRPA has been designed to ensure that the f--sum rule is satisfied. 
It therefore represents an excellent check of the numerical algorithms.
\subsection{Charge density induced by a potential}\label{Sect:results:chargedensity}
The diffusive mode becomes evident when one considers the charged induced
by an external potential as a function of time and distance from its source. The linear response theory 
can only be applied if the potential is weak;  we note that it is well--known that the total particle density surrounding the
potential calculated in linear response may become negative if the potential is strongly repulsive~\cite{Mahan}. Since 
this is obviously a non--physical result, one must be careful in applying linear response theory for 
very strong potentials. In order to use the CF theory as it stands, 
any probe of the 2DEG must be a relatively weak perturbation. Therefore, in this article the potential is 
assumed to result from an electron being placed in a quantum dot of size $l_0$ at a distance $d$
above the plane of the 2DEG. Here $l_0$ and $d$ are both taken to be 5nm for the purposes of illustration; these values are 
physically acceptable but can obviously be varied according to specific device design. The 2DEG density is 
$n_{\text{e}}=10^{15}\text{m}^{-2}$.
Within linear response, the general equation for the charge density induced is~\cite{HLR}
\begin{equation}
J^{\text{ind}}_0({\mathbf{q}},\omega)=-K^{\text{r}}_{00}({\mathbf{q}},\omega) A_0({\mathbf{q}},\omega).
\end{equation}
In this case, the potential due to the charge in the quantum dot is 
\begin{equation}\label{eq:dotpotential}
A_0({\mathbf{q}},t)=-\frac{\theta(t)\,e}{2\epsilon_0 \epsilon_{\text{r}} |{\mathbf{q}}|c} \exp\bigl(-|{\mathbf{q}}|d\bigr) 
\exp\biggl(-\frac{{\mathbf{q}}^2l^2_0}{2}\biggr) 
\end{equation}
Utilizing the dynamic structure factor, and Fourier transforming in time, yields
\begin{multline}
\rho^{\text{ind}}({\mathbf{q}},t)=-\frac{e^2\,\theta(t)}{\epsilon_0 \epsilon_{\text{r}} |{\mathbf{q}}|} 
\exp\bigl(-|{\mathbf{q}}|d\bigr) \exp\biggl(-\frac{{\mathbf{q}}^2l^2_0}{2}\biggr)\\
\int^\infty_0 \hspace{-0.5em}d\omega\, S({\mathbf{q}},\omega) \frac{(1-\cos\,\omega t)}{\omega}.
\end{multline}
Within the diffusive mode approximation, i.e. Eq.~(\ref{eq:diffmode}), the charge induced in the 2DEG builds up with time as 
\begin{equation}
\rho^{\text{ind}}({\mathbf{q}},t)\propto 1-\exp(-t/\tau)
\end{equation}
where the characteristic time $\tau=1/(\eta {\mathbf{q}}^2)$ diverges in the long wavelength limit. Thus the 
system takes a long time to relax to the static density. 
Figures~\ref{fig:cdens1/2}~and~\ref{fig:cdenselec} show the complete particle density as a function 
of the in--plane distance from the dot and the time since the charging of the dot. They clearly show that
the electron system at $B=0$ (Fig.~\ref{fig:cdenselec}) responds much more quickly than does a CF liquid at $B^\star=0$, 
because the abundance of low energy CF excitations means the induced charge continues to fluctuate 
at very long time scales (Fig.~\ref{fig:cdens1/2}).   
\subsection{Quantum dot spectral function}
The dynamic structure is clearly seen to have a large effect on the screening properties of the 2DEG; 
the higher numbers of low energy excitations in the CF liquid at $B^\star=0$ imply that the charge density 
takes a longer time to settle to a constant value than for the electron liquid at $B=0$. The MND theory~\cite{Mahan,Langreth} 
reviewed in Section~\ref{Sect:QDGF} shows that it is also the low energy excitations which determine the shape of the quantum dot 
spectral function, i.e. its density of states. 
\par
Figure~\ref{fig:rho} shows the calculated density of pair excitations created by the dot potential~(\ref{eq:dotpotential}) for the
electrons at $\nu=1/2$, i.e. CFs at $B^\star=0$, and for electrons at $B=0$. 
The figure shows that the density of low energy excitations $\rho(\omega)$
coupling to the potential is much greater for CFs at $B^\star=0$ than for electrons at $B=0$; 
this is a consequence of the relative dynamic structure factors.
The MND theory~\cite{Mahan,Nozieres} shows that it is the low energy behaviour of $\rho(\omega)$ which
is crucial for determining the dot spectral function. 
Analysis of the very low energy spectrum using a log--log plot reveals that while in the case of electron at $B=0$
 $\rho(\omega)\propto\omega$, the density of pair excitations in the CF system at $B^\star=0$ is found to
behave as $\rho(\omega)\propto\omega^{\alpha}$ with $\alpha\approx 0.3$. 
Consequently, the total number of excitations $L$ created in the CF gas at $B^\star=0$ is more strongly divergent than the 
logarithmically--divergent number excited in the electron gas at $B=0$.
This is reflected in the quantum dot spectral functions plotted in Fig.~\ref{fig:spec}.
As usual, the essential comparison is between the spectral functions of the electrons at $\nu=1/2$, i.e. CFs at $B^\star=0$,
and the electron gas at $B=0$. For reference, the spectral function in the absence of the 2DEG
is a delta--function at the zero of energy. The orthogonality catastrophe in the 
electron gas at $B=0$ is manifest as the change of the delta function into a power--law singularity, i.e. it
diverges as an inverse power of the frequency close to threshold. The spectrum at threshold is identically zero. 
This power--law divergence is a direct consequence of the linear behaviour of $\rho(\omega)$ for low
frequencies.~\cite{Mahan} If $\rho$ is linear with $\omega$ up to 
a cut--off $\epsilon_0$, then 
\begin{equation}
\begin{split}
\Gamma(t)=&\,\alpha\int_0^\infty\frac{\text{d}\omega}{\omega} [1-\exp(-i\omega t)]\\ 
\approx&
\begin{cases}
\alpha i\epsilon_0 t & \text{if } \epsilon_0 t \ll 1 \\
\alpha\ln(i\epsilon_0 t) & \text{if } \epsilon_0 t \gg 1 
\end{cases}\\
&\approx \alpha\ln(1+i\epsilon_0 t).\\
\end{split}
\end{equation}
The resulting spectral function is~\cite{Mahan} 
\begin{equation}
\begin{split}
A(\omega)=&\,2\text{Re}\,\int_0^\infty\hspace{-0.5em}\text{d}\omega \exp\,[i\omega t-\Gamma(t)]\\
=&\,\theta(\Omega) \frac{2\sin(\pi \alpha)}{\epsilon_0} \Gamma(1-\alpha) \frac{\exp-\Omega}{\Omega^{1-\alpha}}\\
\end{split}
\end{equation}
where $\omega=\Omega \epsilon_0$. Thus close to threshold, the spectrum goes as $1/\Omega^{1-\alpha}$,
which is a power--law divergence; this is seen in Fig.~\ref{fig:spec}. 
However, the spectral function for $\nu=1/2$ in Fig.~\ref{fig:spec}
is completely different, being suppressed and not divergent 
close to threshold. Since the integral of the spectral function must be conserved~\cite{Mahan}, the low energy weight 
is transferred to higher energies and is manifest as a peak centred near 0.15 V$_{\text{c}}$. This transfer of 
the weight is again a consequence of the far higher number of low energy pair excitations found in the 
CF system at $B^\star=0$ than in the electron gas at $B=0$. This fact is evident from the expansion of the spectral function
in Eq.~(\ref{eq:specexpansion}). The spectral function at energy $\epsilon$ gives the probability of being 
able to create excitations with total energy equal to $\epsilon$. The simplest way of doing this is
to excite one pair with energy $\epsilon$; this is represented by the second term in Eq.~(\ref{eq:specexpansion}). 
However, one can have multi--pair excitations with a total energy $\epsilon$, represented by the 
relevant convolution in the sum, which also contribute to the spectral function. It is precisely these
multipair excitations which have altered the shape from being divergent to suppressed near threshold,
because they have a much higher weight in the CF gas at $B^\star=0$ than in the electron gas at $B=0$. A multipair
excitation contains many, very low--energy pair excitations and sums over all possible combinations. 
Since the CF gas at $B^\star=0$ has more of the very low energy single pair excitations than does the 
electron gas at $B=0$, the contribution will be greater. This interpretation is borne out by calculating the spectrum 
by summing the contributions from the multipair excitations, where it is found that one has to include 
several hundred terms to generate the spectrum.
\par
The shape of the spectral function may be approximately derived using the single diffusive--mode approximation,
taking the quantum dot to have zero size and to lie in the plane of the 2DEG.~\cite{Song} In this case, one finds
\begin{equation}
\rho(\omega)=\biggl(\frac{\omega \Omega}{\pi}\biggr)^{1/2}
\end{equation}
in units of the Coulomb energy, and consequently the spectral function is 
\begin{equation}
A(\omega)=\frac{2\sqrt{\pi}}{\Omega} \biggl(\frac{\Omega}{\omega}\biggr)^{\frac{3}{2}} \exp\biggl(-\frac{
\Omega}{\omega}\biggr)
\end{equation}
where $2\Omega=\pi m\sqrt{m}$.
The density of pair excitations has approximately the correct low energy form, though the cut--off found 
numerically is not present. Likewise, as shown in Fig.~\ref{fig:specapprox}, 
the low energy behavior of the spectral function exhibits a suppression
similar to that found numerically. However, the high energy behaviour, and particularly the positions of 
the peak, do not match. This is not unsuprising because the diffusive mode approximation is only valid
for energies close to threshold.
\section{Odd--denominator states}\label{Sect:ODS}
\subsection{Dynamic structure factor}
The odd--denominator electron states arise from the CF states at integer effective 
filling factors $\nu^\star=p$. The electromagnetic response kernel may be derived,~\cite{Simon1} but
whereas there are explicit expressions for $K^0_{00}, K^0_{01}$ and $K^0_{11}$ for 
the zero effective field case, the expressions for the finite effective field case are sums over
a finite number of filled Landau levels and an infinite number of unfilled levels, which does 
not reduce to a simple expression.~\cite{Simon1} However, these summations may be performed numerically using 
the recursion relations between the Laguerre polynomials. The excitation spectrum is deduced by 
finding the zeros of the denominator of $K_{00}$, which lie on the real axis. The dynamic structure
factor is then simply the residue of $K_{00}$ at this frequency. Consequently, it is numerically 
far more complicated to calculate $S({\mathbf{q}},\omega)$ for the odd--denominator states than it is 
for even--denominator states, particularly with a large number of filled Landau levels. 
Defining the dimensionless density response function as
\begin{gather}
K_{00}({\mathbf{q}},\omega)=\frac{e^2 c^2 m^\star}{\pi \hbar^2} \Pi(x,y)\\
x=\frac{1}{2}{\mathbf{q}}^2 l^{\star 2}_{\text{c}}\hspace{2em}\omega=\omega^\star_{\text{c}} y
\end{gather}
then one can define the dimensionless dynamic structure factor by
\begin{gather}
S({\mathbf{q}},\omega)=\frac{m_{\text{b}}}{\pi \hbar^2} \sigma(x,y)\\
\sigma(x,y)=\frac{m^\star}{m_{\text{b}}} \sum_i \text{Res} \Pi(x,y) \delta(y-y_i(x))
\end{gather}
where the subscript $i$ labels the modes.
\subsection{Pair continuum and cyclotron mode}
The excitation spectrum modes are shown in Figs.~\ref{fig:exc1/3}--\ref{fig:exc10/21}
for the filling factors $\nu=1/3, 3/7, 7/15$ and $10/21$ while the
dynamic structure factor is plotted in the Figs.~\ref{fig:struc1/3}--\ref{fig:struc10/21}.
\par
The modes correspond to CF pair excitations with a given 
momentum, just as in the case of filling factors $\nu=1/2m$. An excitation involves removing a CF
from a given occupied Landau level and putting it in an unoccupied level; this pair excitation
is charge neutral, hence it has a well--defined momentum. The pair carries a dipole moment which is 
proportional to the momentum,~\cite{Simon1,Kallin} and since this affects the overlap of the charge wavefunctions it is 
clear that the interaction energy is affected by the momentum. As ${\mathbf{q}}\rightarrow\infty$ the 
interaction energy tends to zero, thus the pair has an energy equal to the 
effective cyclotron energy multiplied by the number of Landau levels through which the composite 
fermion has been excited.~\cite{Simon1} However, the low ${\mathbf{q}}$ excitations will have energies 
significantly affected by the interactions, giving rise to dispersion. This dispersion is evident 
in all of the modes. The lowest mode is identified with the magnetoroton mode~\cite{Simon1,Chakraborty} 
which has been calculated in the single--mode approximation.~\cite{Chakraborty}
At $\nu=1/3$, there is a deep minimum in the excitation spectrum at 
$ql_{\text{c}}\approx 1$, which is known as the magnetoroton minimum.~\cite{Chakraborty} 
Within the CF model, the lowest excitation mode should be an approximation 
to the magnetoroton mode.~\cite{Simon1} However,  Fig.~\ref{fig:exc1/3} 
shows clearly that the excitation spectrum at $\nu=1/3$ has no minimum. This is an indication
that interactions beyond the MRPA should be included;~\cite{Simon1} an improvement might be to replace the 
non--interacting response kernel ${\mathbf{K}}^0$ by the ladder diagram approximation, which would
generate magnetic excitons in a fashion analogous to those of the $\nu=1$ state considered by 
Kallin and Halperin.~\cite{Kallin} In contrast, the minima are clear at filling factors $\nu=3/7, 7/15$ and $10/21$, as shown
in Figs.~\ref{fig:exc3/7}--\ref{fig:exc10/21}; for these filling factors the MRPA is expected to be a good approximation.   
\par
The weight of these pair excitations is given as the third dimension in
Figs.~\ref{fig:struc1/3}--\ref{fig:struc10/21}. It is clear, particularly from Fig.~\ref{fig:struc10/21}, 
that most of the weight lies within the same region of the momentum and frequency plane as for $\nu=1/2$. 
Note that Fig.~\ref{fig:struc1/3} also shows the cyclotron mode on the 
right of the diagram as a black curve; as in Figs.~\ref{fig:struc1/2}~and~\ref{fig:strucelec}, the residue of this mode
is not shown. The cyclotron mode for the other filling factors lies outside of the plotted range of energies.
\par
The zeros of the denominator of $\Pi(x,y)$ generate the modes mentioned previously. Although 
these are collective, they are identified as the equivalent of the pair continuum excitations at
$\nu=1/2m$. However, this procedure also gives the truly collective cyclotron mode. Figure~\ref{fig:exc1/3}
shows reassuringly that Kohn's theorem is satisfied for $\nu=1/3$, because the long wavelength 
pole is found to be at the bare cyclotron energy. Calculations of the residue along the cyclotron mode
show that the cyclotron mode also contains most of the weight at low ${\mathbf{q}}$, hence it 
exhaust the f--sum rule. This is again a good check of the numerical algorithms.
\subsection{Spectral function of the quantum dot}
The density of pair excitations for the states $\nu=1/3,\ 3/7,\ 7/15$ and $10/21$ have been plotted in 
Fig.~\ref{fig:rho1/3}--\ref{fig:rho10/21} respectively, and there are three essential features to note. 
First, the spectrum has a gap to excitations, therefore one should not expect an orthogonality catastrophe. 
Second, the spectrum is esentially discrete, resulting from the gaps between Landau levels. Near to a turning 
point in the excitation spectrum, the density of states has the one--dimensional inverse
square--root behaviour. Therefore, the density of pair excitations at this energy 
has a singular contribution. Since in a real system the Landau levels would have some broadening, the
spectra shown have been generated by convolution of the bare spectra with a narrow--width 
Lorentzian; 
the actual value of the width is unimportant as long as the spectrum is evaluated at energies 
with a separation much smaller than the linewidth.
Third, as the effective magnetic field is reduced, the envelope of the spectrum tends 
towards the shape of that at $\nu=1/2$. Consequently, one should expect the spectral function
to exhibit similar behaviour.
\par
The quantum dot spectral functions are plotted in Figs.~\ref{fig:spec1/3}--\ref{fig:spec10/21} as 
a function of energy above threshold in units of the Coulomb energy. It should be noted that there is 
a finite weight below the threshold where one would suspect there should be none. This is simply
a result of convolving the bare spectral function, due solely to creation of the pairs, with a narrow
width Lorentzian. Since the excitation spectrum has a gap, the number of pair excitations $L$ created is 
not divergent, and therefore there is no orthogonality catastrophe. Equation~(\ref{eq:specexpansion}) shows that there will be 
a finite probability of creating no excitations at all, which is manifest as a delta--function peak 
at threshold with weight $\exp(-L)$. Since there will be a limiting resolution in any experiment 
designed to measure the quantum dot density of states, it is appropriate to apply a broadening, which in this
case is of order 0.1 meV. This is comparable to the effective cyclotron energy for the given density 
at $\nu=10/21$, thus the detail in the spectrum at this filling factor is poorly resolved, as seen in Fig.~\ref{fig:spec10/21}.
However, the structure in the spectra for the states $\nu=1/3,\ 3/7$ and $7/15$ is resolved; in 
each of the Figs.~\ref{fig:spec1/3}--\ref{fig:spec7/15},
one may clearly see an oscillation with a period equal to the effective cyclotron energy. This result is simply a consequence of the
gaps in the excitation spectra.
\par
The evolution of the spectrum as the real magnetic field, and hence the effective field, is reduced 
(Figs.~\ref{fig:spec1/3}--\ref{fig:spec10/21}) is easily explained by the change in density of pair excitations 
(Figs.~\ref{fig:rho1/3}--\ref{fig:rho10/21}). For a system with a gap, there is no 
orthogonality catastrophe, therefore there is a finite spectral weight at threshold. The expansion
of the spectral function Eq.~(\ref{eq:specexpansion}) shows that this weight goes as $\exp(-L)$ where $L$
is the mean number of excitations which are created. (A diverging $L$ is indicative of a catastrophe).
As the field is reduced, the energy gap decreases in size and the average number of excitations increases,
hence the weight at threshold decreases. This feature is obvious from 
Figs.~\ref{fig:spec1/3}--\ref{fig:spec10/21}. However what is perhaps more interesting is that as the field is 
reduced, the spectrum quickly develops a low energy suppression and a high energy peak, reminiscent of that at 
$\nu=1/2$. This is attributable to the change in the density of pair excitations with field, which as already mentioned
develops an envelope similar to $\rho(\omega)$ at $\nu=1/2$. 
However, the spectra also show the oscillations with period equal to the effective cyclotron energy 
superimposed on this envelope. 
\par
A comparison with the results for ordinary electrons at the equivalent filling factors of 
$\nu=1, 3, 7$ and 10 is, at this point, informative. Within the same level of approximation as 
used for the CFs, i.e. the (M)RPA, the spectra have oscillations characteristic of the cyclotron energy 
similar to the case above of composite fermions.~\cite{Uenoyama2,Westfahl,Uenoyama1} However, where 
the behaviour of the CF and electron spectra differ is in their respective low--field regimes. 
The threshold peak for the electrons does fall in value as the field is reduced, being the largest 
feature in the spectrum at $\nu=1$ and tending to zero as the field 
is switched off, hence giving the orthogonality catastrophe viewed in Fig.~\ref{fig:spec}. 
However, it is still essentially peaked close to threshold, unlike in the CF case where a
shift in weight is observed as the effective field is reduced from $\nu=1/3$ to $\nu=1/2$. 
\par
It should be noted that to discuss the $\nu=1$ state in such an approximation is at best naive and at worst incorrect. This is in
precisely the range of filling factors where the electron--electron interaction causes the correlations associated with the FQHR, 
so the electron--electron interaction cannot be treated as a weak perturbation. 
However, this is analogous to employing the MRPA for CFs at $\nu^\star=1$,
and therefore, as a method of comparing electrons with CFs, this level of approximation is valid.
As stated previously, a better approximation could be to extend the works of Kallin and Halperin~\cite{Kallin} by
using both the Chern--Simons and Coulomb interactions. This point is not pursued here.
\section{Proposed experimental probe of the dynamic response of the composite fermion gas}\label{Sect:Tunneling}
In the previous section, it was shown how the neutral excitations of the 2DEG may be 
probed by placing an electron into an quantum dot situated in the vicinity of the 2DEG. The density of states of the 
quantum dot reflects the neutral excitation spectrum of the 2DEG.
However, the quantum dot spectral function is a purely theoretical construct, therefore it needs to be related to a quantity 
which is experimentally measurable. In metals, inverse photoemission spectroscopy (IPS)~\cite{Mahan,Fuggle} may be used to 
determine the neutral excitation spectrum of the conduction band because the 
density of states of a core atomic potential in a metal can be measured directly as the IPS spectrum.
However, it would not be easy to control the proximity of the core hole to the 2DEG, so it is unlikely to be 
a useful method for measuring specifically the 2DEG density of states.
\par
One possible scheme for bringing a local quantum dot close to the 2DEG is to allow electrons from a lead 
to tunnel into a single energy level in an impurity state or quantum dot specially grown in the
vicinity of the 2DEG. An electron in the dot will interact with the 2DEG and, as a consequence, the dot's density
of states will broaden reflecting the creation of pair excitations. By keeping the energy of the 
tunneling electron fixed, one may vary the energy of the dot by applying a gate voltage which brings
different regions of the dot density of states into resonance with the tunneling electron. The 
current measured as a function of the gate voltage should be proportional to the dot spectral function at that energy.
\subsection{Tunneling experiment}\label{Sect:tunnel:expt}
Figure~\ref{fig:junction} provides a schematic illustration of the proposed
experiment. The lower portion shows the geometry, while the upper portion
describes the energetics of the tunneling process. There is one feature which 
needs particular explanation, and this is the presence of two quantum dots. In general, the tunneling 
process takes an electron from the source lead. This electron can have a range of energies up to 
the Fermi energy of the lead, and consequently without the presence of dot A the current would be 
proportional to the convolution of the density of states of the lead and that of dot B. Hence the 
$I$--$V$ profile in this case is not a direct measure of dot B's spectral function.
The purpose of dot A is therefore to act as an electron monochromator, because only 
source electrons with energy $\epsilon$ can resonantly tunnel to A. An
electron in A will resonantly
tunnel to quantum dot B only if there are states available with energy
$\epsilon$. In the absence of the
2DEG the density of states of B is a $\delta-$function at energy
$\epsilon_0$, and tunneling occurs
only when  $\epsilon=\epsilon_0$. However, the presence of the 2DEG means
that the density of states is 
asymmetrically broadened to higher energies due to the neutral excitations
in the 
2DEG induced by the filled dot -- this implies that tunneling can only occur 
if $\epsilon>\epsilon_0$. The electron can then tunnel from dot B 
into the drain lead to be measured as a current, determined by the
tunneling rates 
$\gamma_{\mathrm{a}},\ \Gamma$, and $\gamma_{\mathrm{b}}$.
The current $I$ is measured as a function of the 
gate voltage $V$ controlling the difference between the 
energy $\epsilon$ of the injected electron and the energy $\epsilon_0$
of dot B.
This resonant tunneling is similar to IPS with a zero energy photon, the
analog of the IPS
spectrum being the tunneling I-V characteristic; the threshold $V=V_0$ in
this case is such that 
$\epsilon_0=\epsilon$. Generally, $\epsilon-\epsilon_0=e(V-V_0)$, hence
the creation of excitations implies that the spectrum is non-zero for
$V>V_0$. 
\par 
Once an electron has tunneled into dot B it must reside there for a time
greater than the response time
of the 2DEG. This implies that the electron tunnels out with rate
$\gamma_{\mathrm{b}}$ less than the desired resolution, which is typically the 
CF Landau-level spacing for 
a filling factor close to $\nu=1/2$. The other two rates $\Gamma$ and $\gamma_{\mathrm{a}}$ are determined by the 
following simple argument which will be justified later. The average current can be written as
\begin{equation}
I=-\frac{e}{T}
\end{equation}
where $T$ is the total time taken to tunnel from source to drain. In terms of the tunneling rates we have for sequential tunneling
\begin{equation}
T\sim\frac{1}{\gamma_{\mathrm{a}}}+\frac{1}{\Gamma}+\frac{1}{\gamma_{\mathrm{b}}}.
\end{equation}
In order for the tunneling current to reflect only the density of states of dot B, the conditions
\begin{equation}\label{eq:conditions}
\gamma_{\mathrm{a}} \sim \gamma_{\mathrm{b}} \equiv \gamma \qquad \Gamma \ll \gamma
\qquad \gamma \leq \omega_{\mathrm{c}}^\star
\end{equation} 
are chosen, in which case 
\begin{equation}\label{eq:Current1}
I=-e\Gamma.
\end{equation}
The rate of tunneling from A to B is designed to be slow, while that from B to the drain is large. 
Tunneling from the source lead into dot A is arranged to be so fast that A may be considered to be
essentially always full. Therefore dot A may be regarded as though it is a lead with a Lorentzian density of states.
Tunneling then occurs from this lead to the drain lead via a localized state in dot B. Consequently one may
use the tunneling Hamiltonian approach~\cite{Matveev,Larkin,Datta} with A as the first lead and the drain as the second lead;
resonant tunneling occurs from the localized state B.  Since the tunneling rate out of B is very fast, 
it is the rate of tunneling \emph{into} the dot which is of relevance when determining the current. 
The current is thus determined by the rate of change of the occupancy of dot B. The Hamiltonian may be written
\begin{equation}
\begin{split}
H=&\bar{H}_0+H_{\text{T}}\\
\bar{H}_0=&H_{2\text{DEG}}+\epsilon_{\text{a}} a^\dag a + \epsilon_{\text{b}} b^\dag b + V b^\dag b \\
V=&\sum_{\mathbf{q}} V(\mathbf{q}) \rho(\mathbf{q})\\
H_{\text{T}}=& J a^\dag b + J^\star b^\dag a\\
\end{split}
\end{equation}
where dot A has energy $\epsilon_{\text{a}}\equiv\hbar\omega_{\text{a}}$ with operators $a$ and $a^\dag$, 
while dot B has energy $\epsilon_{\text{b}}\equiv\hbar\omega_{\text{b}}$ with operators $b$ and $b^\dag$.
The 2DEG is described by the Hamiltonian $H_{\text{2DEG}}$, $V$ is the interaction of the dot with
the 2DEG, and the electron may hop between the dots with matrix element $J$. 
The current is simply
\begin{equation}
I=-e\frac{d}{d t} \biggl[ \bra{\psi(t)} b^\dag b \ket{\psi(t)} \biggr]
\end{equation}
where the state of the system must satisfy the time--dependent Schrodinger equation  
\begin{equation}
i\hbar\frac{\partial}{\partial t} \ket{\psi(t)} = H\ket{\psi(t)}.
\end{equation}
Using this, and going from the Schrodinger to the Heisenberg representation, the current is 
\begin{equation}
\begin{split}
I=&-\frac{ie}{\hbar}\bra{\psi(0)} [H,b^\dag_{\text{H}}(t) b_{\text{H}}(t)] \ket{\psi(0)}\\
=&\frac{2e}{\hbar} \text{Im}\,\bra{\psi(0)} J a^\dag_{\text{H}}(t) b_{\text{H}}(t) \ket{\psi(0)}\\
\end{split}
\end{equation}
where an operator $A$ becomes 
\begin{equation}
A_{\text{H}}(t)=\exp(iHt/\hbar)\,A\,\exp(-iHt/\hbar)
\end{equation}
 in the Heisenberg representation.
Now the assumption is made that the state $\ket{\psi(0)}$ is adiabatically evolved from the `unperturbed'
state $\ket{\bar{0}}$ which is an eigenstate of $\bar{H}_0$ as in 
\begin{equation}
\ket{\psi(0)}=S_{\text{T}}(0,-\infty) \ket{\bar{0}}
\end{equation}
where 
\begin{equation}
S_{\text{T}}(t,t')=U_{\text{T}}(t) U^\dag_{\text{T}}(t')
\end{equation}
and 
\begin{equation}
U_{\text{T}}(t)=\exp(i\bar{H}_0 t/\hbar)\,\exp(iH t/\hbar).
\end{equation}
In this case, the current may be written
in terms of the interaction representation operators 
\begin{equation}
\bar{A}(t)=\exp(i\bar{H}_0t/\hbar)\,A\,
\exp(-i\bar{H}_0t/\hbar)
\end{equation}
as
\begin{equation}
\begin{split}
I=&\frac{2e}{\hbar} \text{Im}\,J\bra{\bar{0}} S^\dag_{\text{T}}(t,-\infty) \bar{a}^\dag(t) \bar{b}(t)
S(t,-\infty) \ket{\bar{0}}\\
\approx&\frac{2e}{\hbar} \text{Im}\,J\bra{\bar{0}}\biggl[1+\frac{i}{\hbar}\int^t_{-\infty}\hspace{-1em}dt_1 \bar{H}_
{\text{T}}(t_1)\biggr]\times\\
&\hspace{4em}\bar{a}^\dag(t) \bar{b}(t)\biggl[1-\frac{i}{\hbar}\int^t_{-\infty}\hspace{-1em}dt_1 \bar{H}_
{\text{T}}(t_1)\biggr]\ket{\bar{0}}\\
\approx&-\frac{2e}{\hbar^2} |J|^2 \text{Re}\,\int^t_{-\infty} 
\hspace{-1em}dt_1 \bra{\bar{0}} \bar{a}^\dag(t) \bar{b}(t)
\bar{b}^\dag(t_1) \bar{a}(t_1)\ket{\bar{0}}\\
\end{split}
\end{equation}
to first order in the perturbation $H_{\text{T}}$. This may be rewritten in terms of the spectral 
function $A_{\text{B}}(\omega)$ of dot B  as 
\begin{equation}
I=-\frac{e |J|^2}{\hbar} A_{\text{B}}(\omega_{\text{a}}) 
\end{equation}
from which it is clear that the current is a direct measure of the density of states. 
The density of states of dot A is broadened into a Lorentzian following the tunneling process from the source, the width $\gamma$ 
of this Lorentzian being thought of as the instrumental resolution. Thus the current is 
\begin{equation}\label{eq:Current2}
I=-\frac{2 e |J|^2}{\hbar^2}\,\text{Re}\,\int_0^\infty\hspace{-0.5em}
dt \exp\bigl[i(\omega_{\text{a}}-\bar{\omega}_{\text{b}}+i\gamma)t-F(t)\bigr],
\end{equation}
where 
\begin{equation}
F(t)=\int_0^\infty\hspace{-0.5em}d\omega\,\frac{(1-e^{-i\omega t})}{\omega^2}\,\rho(\omega)
\end{equation}
and
\begin{equation}
\rho(\omega)=\frac{1}{\hbar} \sum_{\mathbf{q}} |V({\mathbf{q}})|^2 S({\mathbf{q}},\omega).
\end{equation}
The energy difference between dots A and B is controlled by a gate voltage, $V$, i.e.
\begin{equation}
\hbar(\omega_{\text{a}}-\bar{\omega}_{\text{b}})=e(V-V_0);
\end{equation}
$V_0$ is the threshold voltage. Equating Eq.~(\ref{eq:Current1}) with Eq.~(\ref{eq:Current2}) implies that 
\begin{equation}
\Gamma=\frac{2|J|^2}{\hbar^2}\,\text{Re}\,\int_0^\infty\hspace{-0.5em} 
dt \exp\bigl[i(\omega_{\text{a}}-\bar{\omega}_{\text{b}}+i\gamma)t-
F(t)\bigr].
\end{equation}
The maximum magntitude of $\hbar\Gamma$ in the absence of the 2DEG is thus 
\begin{equation}\label{eq:maxcurrent}
\hbar\Gamma_0=\frac{2|J|^2}{\hbar\gamma}.
\end{equation}
If the limiting resolution is the Landau level separation at $\nu=10/21$, then from Fig.~\ref{fig:spec10/21} one
observes that a suitable value is $\hbar\gamma\approx0.1$ meV. Satisfying conditions~(\ref{eq:conditions}) therefore implies a value 
 $\hbar\Gamma_0\approx 5\ \mu$eV. Hence Eq.~(\ref{eq:maxcurrent}) suggests a value of $|J|\approx 3.6\times10^{-24}$eV.
\subsection{Junction design}
In order that the above theory is a good description of the tunneling process, and therefore that the 
experiment achieves the goal of measuring the spectral function of dot B, there are several criteria which must be satisfied in the 
design of the junction.
\begin{enumerate}
\item The source and drain leads must be taken to be either one-- or three--dimensional, so that the 
effect on them due to the magnetic field may be considered a weak perturbation.
\item The quantum dot must be far enough away from the 2DEG so as to allow the linear response
functions to be used, but not so far as to be too weak a perturbation to couple to the 2DEG. 
This is controlled by the separation $d$. The potential $V(\mathbf{q})$ scales as $\exp-|{\mathbf{q}}|d$,
hence if $d$ is too large, the quantum dot does not couple effectively to the low energy excitations. 
In particular, the oscillations of the spectral function due to the Landau level 
structure would not be seen.
\item The quantum dot must also be relatively small in size otherwise the form factor 
$\exp-\frac{1}{2}{\mathbf{q}}^2l^2_0$ destroys the structure of interest.
\item Related to the point above, the level separation in the dot must be much larger than the 
range of the density of states induced by the 2DEG, which again requires a small dot. 
If this is not the case, the quantum dot can no longer be approximated as a single level system, 
and the MND techniques are not applicable. 
\item The quantum dot may also influence and be influenced by the drain electrons, which means that 
the signal will be a convolution of the effect due to the 2DEG with that due to the drain. This problem
is minimized if the drain is kept sufficiently far enough from B, since the potential is short ranged. 
\end{enumerate}
There is also the question of the barrier dimensions.  The tunneling matrix element
$J$ may be approximately derived for the junction shown in Fig.~\ref{fig:barrier}, 
which has two dots either side of the potential barrier.~\cite{Larkin,Datta} 
For quantum dots with size $l_0\approx 50$ \AA, the value $|J|\approx 3.6\times10^{-24}$eV determined above
requires a junction with $a\approx100$ \AA, $b\approx250$ \AA, 
$c\approx150$ \AA\ and $V_{\text{B}}\approx
0.8$ eV; this results in a current $I_{\text{max}}\sim 2$ nA.
\subsection{I--V characteristics}
Figures~\ref{fig:current1/2}--\ref{fig:current7/15} show the I--V characteristics for the junction with
the above conditions satisfied, and they clearly reflect the quantum dot spectral functions shown in the 
previous section (Figs.~\ref{fig:spec},~and~\ref{fig:spec1/3}--\ref{fig:spec7/15}). The instrumental 
resolution represented by $\gamma$ is responsible for both the non--zero current below threshold 
and also for smearing out the oscillations. 
\par
As detailed in the previous section, the spectral function at a given filling factor
reflects the density--density excitation spectrum of the 2DEG state. The I--V characteristic can therefore
be used as a test of the CF theory of the FQHR since one should be able to
\begin{enumerate}
\item observe the evolution of the excitation spectrum from gapless to gapped
\item measure the CF effective mass from the period of the oscillations in the spectrum.
\end{enumerate}
\section{Conclusion}\label{Sect:Conclusions}
In this paper we have calculated the response of a 2DEG to the sudden appearance of a charged probe within the Chern--Simons
composite fermion model. For the purposes of this paper 
the probe was assumed to be an electron in a quantum dot situated in a plane a distance $d$ above the
plane of the 2DEG. However, we emphasize that the response of the 2DEG calculated in Secs.~\ref{Sect:QDGF}--\ref{Sect:ODS}
is completely general. The formalism is based on the theory developed by MND in connection with the x--ray edge problem, 
and is it shown that the response of the 2DEG to the probe is encapsulated in the spectral function of the probe. This is found to be
a function of the dynamic structure factor of the 2DEG. We used the MRPA to deduce $S({\mathbf{q}},\omega)$ for  CFs with
an effective mass scaling as the square-root of the magnetic field at filling factors 
 $\nu=1/2$, 10/21, 7/15, 3/7 and 1/3. We also calculated the $S({\mathbf{q}},\omega)$
for electrons in zero magnetic field analogous to CFs in zero effective field, and found that there are more low ${\mathbf{q}}$ and
 $\omega$ excitations in the CF gas at $B^\star=0$ than there are in the electron gas at $B=0$; 
this is a consequence of the extra scattering in the CF gas 
resulting from the Chern--Simons potential which appears in the CF Hamiltonian but not in the electron Hamiltonian.
These excitations are found to greatly affect the screening properties of the system. We deduced the charge induced by a weak probe 
in the CF and electron gases at zero effective and real fields respectively within linear response and found that the greater number of
excitations in the CF gas at $B^\star=0$ causes the charge density of the CFs to fluctuate for a much longer time than it 
does in the electron gas at $B=0$; this is in agreement with the observation that the charge density relaxation time calculated 
using the single diffusive mode approximation diverges in the long--wavelength limit.
\par
We then determined the spectral function of the probe at the various filling factors. This depends on the density of pair excitations
created by the probe $\rho(\omega)$. According to the MND theory, an orthogonality catastrophe is predicted if 
the excitation spectrum is gapless and the low--energy behaviour of $\rho(\omega)$ is weaker than linear. This is because the mean
number of excitations created diverges causing the Debye--Waller factor to tend to zero. The electron gas in zero magnetic field
is shown to exhibit an orthogonality catastrophe resulting in a the well--known power--law divergence of the probe spectral 
function at threshold, because $\rho(\omega)\propto\omega$ for low $\omega$ . The CF gas in zero effective field
is also gapless but, due to the larger number of excitations, one finds the divergence of the mean number of excitations created by 
the probe in the CF gas is greater, thus the orthogonality catastrophe is more severe. It is found that the probe spectral function is now
heavily suppressed at threshold. However, a CF gas in finite effective field has a gap to excitations, consequently there is no
orthogonality catastrophe. The probe spectral function therefore has a $\delta$--peak at threshold with a non--zero weight 
equal to the Debye--Waller factor. There is an oscillation in the spectral function which has a period equal to the effective 
cyclotron energy of the CFs; the envelope of the oscillations becomes more like the shape of the spectral function in zero effective
field as the effective field is reduced.
\par 
Our work shows that by measuring the spectral function of the probe one could determine the effective mass of the CFs
as a function of field. To this end, we proposed an experiment which measures the spectral function of the probe directly. The
probe is taken to be an electron in a quantum dot placed a distance $d$ above the plane of the 2DEG. The sudden appearance of the
electron in the dot is accomplished by a tunneling process. A two--dot device is proposed with the first dot acting as an 
energy filter and the second dot as the probe. An electron resonantly tunnels from a source lead into the first dot, then tunnels into the
probe dot whereupon it interacts with the 2DEG. After a time longer than the response time of the 2DEG the electron 
tunnels out of the probe to a drain lead. The interaction broadens the density of states of the probe dot so that there 
is a finite tunneling probability proportional to the density of states at the energy of the monochromator dot. The difference between
the energies of the two dots is controlled by the voltage difference between the dots, thus this voltage alters the density of states
of the probe dot at the energy of the monochromator dot. Since the current is proportional to this density of states, a plot 
of current through the device against gate voltage gives the spectral function of the probe dot resulting from the interaction of the
dot with the 2DEG. Therefore, the I--V characteristic of the device is the spectral function of the dot, and it may be used
to examine the excitation spectrum of the 2DEG as a function of magnetic field. In particular, we propose that the period of
oscillations in the I--V profile at odd--denominator filling factors may be used to determine the effective mass of the CFs. 
\section*{Acknowledgements}
We thank B. I. Halperin, S. H. Simon and T. Portengen for useful discussions. 
We acknowledge the financial support of EPSRC through a Studentship (D.J.T.L) and
EPSRC grant No. GR/K 15619.

\end{multicols}
\bigskip
\begin{center}
\textbf{Figure Captions}
\end{center}
\begin{figure}
\caption[Structure factor for $\nu=1/2$.]
{Dimensionless dynamic structure factor $\sigma$ of electrons at filling factor $\nu=1/2$ within the composite fermion
model. The CFs experience an effective magnetic field $B^\star=0$. The momentum is in units of $1/l_{\text{c}}$ and 
the energy is in units of the Coulomb energy. The solid line shows the collective mode dispersion.}
\label{fig:struc1/2}
\end{figure}
\begin{figure}
\caption[Structure factor of electrons.]
{Dimensionless dynamic structure factor $\sigma$ of electrons at zero $B$--field. The momentum is in units of $k_{\text{f}}$ 
and the energy is in units of the Coulomb energy. The solid line shows the collective mode dispersion.}
\label{fig:strucelec}
\end{figure}
\begin{figure}
\caption[Charge density at $\nu=1/2$.]
{Charge density surrounding the charged quantum dot, for an electron gas at filling factor $\nu=1/2$ (CFs in $B^\star=0$)
as a function of the in--plane distance in nanometres from the dot, and the time in picoseconds after the dot was charged.}
\label{fig:cdens1/2}
\end{figure}
\begin{figure}
\caption[Charge density for electrons.]
{Charge density surrounding the charged quantum dot, for an electron gas at zero magnetic field as a function of the in--plane 
distance in nanometres from the dot and the time in picoseconds after the dot was charged.}
\label{fig:cdenselec}
\end{figure}
\begin{figure}
\caption[Density of pair excitations $\rho(\omega)$ for an electron gas at $\nu=1/2$ (CFs at $B^\star=0$) and for electrons
in zero field.]
{Density of pair excitations $\rho(\omega)$ for an electron gas at $\nu=1/2$ (CFs at $B^\star=0$) and for electrons
in zero field. The energy is in units of the Coulomb energy.}
\label{fig:rho}
\end{figure}
\begin{figure}
\caption[Spectral function $A(\omega)$ for an electron gas at $\nu=1/2$ (CFs at $B^\star=0$) and for electrons
in zero field.]
{Quantum dot spectral function $A(\omega)$ for an electron gas at $\nu=1/2$ (CFs at $B^\star=0$) and for electrons
in zero field. The energy is measured from threshold in units of the Coulomb energy.}
\label{fig:spec}
\end{figure}
\begin{figure}
\caption[Approximation to $A(\omega)\text{ at }\nu=1/2$.] 
{Quantum dot spectral function $A(\omega)$ for electrons at $\nu=1/2$ in the diffusive mode approximation. 
The energy is in units of the Coulomb energy.}
\label{fig:specapprox}
\end{figure}
\begin{figure}
\caption[Excitation spectrum at $\nu=1/3$.]
{Excitation spectrum at filling factor $\nu=1/3$. The momentum is in
units of $1/l_{\text{c}}$ and the energy is in units of the Coulomb energy.}
\label{fig:exc1/3}
\end{figure}
\begin{figure}
\caption[Excitation spectrum at $\nu=3/7$.]
{Excitation spectrum at filling factor $\nu=3/7$. The momentum is in
units of $1/l_{\text{c}}$ and the energy is in units of the Coulomb energy.}
\label{fig:exc3/7}
\end{figure}
\begin{figure}
\caption[Excitation spectrum at $\nu=7/15$.]
{Excitation spectrum at filling factor $\nu=7/15$. The momentum is in
units of $1/l_{\text{c}}$ and the energy is in units of the Coulomb energy.}
\label{fig:exc7/15}
\end{figure}
\begin{figure}
\caption[Excitation spectrum at $\nu=10/21$.]
{Excitation spectrum at filling factor $\nu=10/21$. The momentum is in
units of $1/l_{\text{c}}$ and the energy is in units of the Coulomb energy.}
\label{fig:exc10/21}
\end{figure}
\begin{figure}
\caption[Structure factor at $\nu=1/3$.]
{Dimensionless dynamic structure factor $\sigma$ at filling factor $\nu=1/3$. The momentum is in
units of $1/l_{\text{c}}$ and the energy is in units of the Coulomb energy.}
\label{fig:struc1/3}
\end{figure}
\begin{figure}
\caption[Structure factor at $\nu=3/7$.]
{Dimensionless dynamic structure factor $\sigma$ at filling factor $\nu=3/7$. The momentum is in
units of $1/l_{\text{c}}$ and the energy is in units of the Coulomb energy.}
\label{fig:struc3/7}
\end{figure}
\begin{figure}
\caption[Structure factor at $\nu=7/15$.]
{Dimensionless dynamic structure factor $\sigma$ at filling factor $\nu=7/15$. The momentum is in
units of $1/l_{\text{c}}$ and the energy is in units of the Coulomb energy.}
\label{fig:struc7/15}
\end{figure}
\begin{figure}
\caption[Structure factor at $\nu=10/21$.]
{Dimensionless dynamic structure factor at filling factor $\nu=10/21$. The momentum is in
units of $1/l_{\text{c}}$ and the energy is in units of the Coulomb energy.}
\label{fig:struc10/21}
\end{figure}
\begin{figure}
\caption[$\rho(\omega)$ at $\nu=1/3$.]
{Density of pair excitations $\rho(\omega)$ at $\nu=1/3$. The energy is in units of the Coulomb energy.}
\label{fig:rho1/3}
\end{figure}
\begin{figure}
\caption[$\rho(\omega)$ at $\nu=3/7$.]
{Density of pair excitations $\rho(\omega)$ at $\nu=3/7$. The energy is in units of the Coulomb energy.}
\label{fig:rho3/7}
\end{figure}
\begin{figure}
\caption[$\rho(\omega)$ at $\nu=7/15$.]
{Density of pair excitations $\rho(\omega)$ at $\nu=7/15$. The energy is in units of the Coulomb energy.}
\label{fig:rho7/15}
\end{figure}
\begin{figure}
\caption[$\rho(\omega)$ at $\nu=10/21$.]
{Density of pair excitations $\rho(\omega)$ at $\nu=10/21$. The energy is in units of the Coulomb energy.}
\label{fig:rho10/21}
\end{figure}
\begin{figure}
\caption[$A(\omega)$ at $\nu=1/3$.]
{Spectral function $A(\omega)$ at $\nu=1/3$. The energy is measured from threshold in units of the Coulomb energy.}
\label{fig:spec1/3}
\end{figure}
\begin{figure}
\caption[$A(\omega)$ at $\nu=3/7$.]
{Spectral function $A(\omega)$ at $\nu=3/7$. The energy is measured from threshold in units of the Coulomb energy.}
\label{fig:spec3/7}
\end{figure}
\begin{figure}
\caption[$A(\omega)$ at $\nu=7/15$.]
{Spectral function $A(\omega)$ at $\nu=7/15$. The energy is measured from threshold in units of the Coulomb energy.}
\label{fig:spec7/15}
\end{figure}
\begin{figure}
\caption[$A(\omega)$ at $\nu=10/21$.]
{Spectral function $A(\omega)$ at $\nu=10/21$. The energy is measured from threshold in units of the Coulomb energy.}
\label{fig:spec10/21}
\end{figure}
\begin{figure}
\caption[Diagram of the tunneling junction.]
{Schematic diagram of the junction. A gate voltage $V$ is applied to alter the
energy of dot B with respect to A.
The source-drain voltage is kept fixed. The plane of the 2DEG is
perpendicular to the page.}
\label{fig:junction}
\end{figure}
\begin{figure}
\caption[Dimensions of the central barrier of the junction.]
{Dimensions of the central barrier of the junction.}
\label{fig:barrier}
\end{figure}
\begin{figure}
\caption[I--V trace at $\nu=1/2$.]
{Current--voltage characteristic at $\nu=1/2$.}
\label{fig:current1/2}
\end{figure}
\begin{figure}
\caption[I--V trace at $\nu=1/3$.]
{Current--voltage characteristic at $\nu=1/3$.}
\label{fig:current1/3}
\end{figure}
\begin{figure}
\caption[I--V trace at $\nu=3/7$.]
{Current--voltage characteristic at $\nu=3/7$.}
\label{fig:current3/7}
\end{figure}
\begin{figure}
\caption[I--V trace at $\nu=7/15$.]
{Current--voltage characteristic at $\nu=7/15$.}
\label{fig:current7/15}
\end{figure}
\end{document}